\documentclass[submission,copyright,creativecommons]{eptcs}
\providecommand{\event}{ICLP 2024} 

\usepackage{iftex}
\usepackage{listings}
\usepackage{xcolor}

\definecolor{codegreen}{rgb}{0,0.6,0}
\definecolor{codegray}{rgb}{0.5,0.5,0.5}
\definecolor{codepurple}{rgb}{0.58,0,0.82}
\definecolor{backcolour}{rgb}{0.95,0.95,0.92}

\lstdefinestyle{mystyle}{
    backgroundcolor=\color{backcolour},   
    commentstyle=\color{codegreen},
    keywordstyle=\color{magenta},
    numberstyle=\tiny\color{codegray},
    stringstyle=\color{codepurple},
    basicstyle=\ttfamily\footnotesize,
    breakatwhitespace=false,         
    breaklines=true,                 
    captionpos=b,                    
    keepspaces=false,                 
    numbers=left,                    
    numbersep=5pt,                  
    showspaces=false,                
    showstringspaces=false,
    showtabs=false,                  
    tabsize=2
}

\ifpdf
  \usepackage{underscore}         
  \usepackage[T1]{fontenc}        
\else
  \usepackage{breakurl}           
\fi

\usepackage{amsthm}

\newtheorem{theorem}{Theorem}
\newtheorem{lemma}{Lemma}

\title{A Coq Formalization of Unification Modulo Exclusive-Or}
\author{Yichi Xu 
\institute{Worcester Polytechnic Institute\\
Massachusetts, USA}
\email{yxu10@wpi.edu}
\and Daniel J. Dougherty 
\institute{Worcester Polytechnic Institute\\
Massachusetts, USA}
\email{dd@wpi.edu}
\and Rose Bohrer
\institute{Worcester Polytechnic Institute\\
Massachusetts, USA}
\email{rbohrer@wpi.edu}
}
\def\titlerunning{A Coq Formalization of Unification Modulo Exclusive-Or}
\def\authorrunning{Y. Xu, D. Dougherty, and R. Bohrer}
\begin{document}
\maketitle

\begin{abstract}
\textit{Equational Unification} is a critical problem in many areas such as
  automated theorem proving and security protocol analysis. In this
  paper, we focus on XOR-Unification, that is, unification modulo the
  theory of exclusive-or.  This theory contains an operator with the
  properties Associativity, Commutativity, Nilpotency, and the
  presence of an identity. In the proof assistant Coq, we 
  implement an algorithm that solves XOR unification problems, whose design was inspired by Liu and Lynch,
  and prove it sound, complete, and terminating. Using Coq's
  code extraction capability we obtain an implementation in the
  programming language OCaml.
\end{abstract}

\section{Introduction}
Unification is a fundamental concept used across various domains such as logic
programming, type systems, and constraint solving. In logic programming, it enables
pattern matching and logical inference, crucial for problem-solving in languages 
like Prolog \cite{clocksin2003programming}. Within type systems, such as those in Haskell and Scala \cite{odersky2016programming}, unification
supports type inference, allowing compilers to ensure type safety and catch type errors
at compile-time rather than runtime. Additionally, in constraint solving, it helps 
manage and resolve variable constraints, essential for applications in scheduling 
and planning. 
This work, however, is motivated by applications of unification to security protocol analysis.
A common way to analyze protocols is to perform syntactic unification with 
the protocol rules to explore some space of reachable states. 
If an ``attack'' state is reachable from the initial state then an attack 
exists and the protocol is flawed. 

However, the limitation of using syntactic unification to analyze protocols is that it 
only captures the case when terms, representing messages, can be made exactly the same, which
in many protocols is not enough. 
For example, the Vernam cipher and cipher-block chaining mode for block
ciphers rely on exclusive-or (XOR)~\cite{menezes2018handbook}.
There exists  protocols which seem secured if XOR is left uninterpreted, but whose flaws are revealed when XOR is an interpreted operator.
For example, the original version of Bull’s recursive authentication protocol was formally proved correct in the Dolev-Yao model, but this XOR-based protocol was vulnerable to an attack that exploited the self-cancellation property~\cite{10.1016/S0020-0190(97)00180-4}.
XOR Unification is important because it enables a more accurate analysis of XOR-based protocols.
Because unification is a key ingredient of logic programming and because logic programming has established applications to security analysis~\cite{zech2019knowledge}, we believe research about XOR Unification may help enable logic programming-based analyses of XOR-based protocols in the long term.

In this paper, we adopt a modified version of the algorithm developed
by Liu and Lynch~\cite{liu2011efficient}, then implement it and
prove it correct in Coq. 
This work is important because it increases our confidence in Liu and Lynch's work. 
The primary reason for not adopting  the full algorithm was time constraints. Consequently, we decided to exclude uninterpreted  functions and homomorphic function from our implementation. 
This decision was  based on the understanding that both uninterpreted and homomorphic functions can be treated as standard terms under specific constraints. 
For uninterpreted  functions, we can compare their function symbols and conduct 
syntactic unification within the functions. For homomorphic functions, we 
can reduce them to their normal form before proceeding with syntactic comparison. 

\section{Related Work}

\emph{Syntactic} unification is unification modulo the empty
equational theory. There are many algorithms for syntactic unification, but there are
only a few which have been verified and formalized. The earliest formalization
is the algorithm from Manna and Waldinger \cite{manna1983deductive}, which was proved by Paulson \cite{paulson1985verifying} using LCF (Logic for Computable Functions). 
This formalization is used as a basis for later research by Sternagel and Thiemann 
\cite{FirstOrderTermsAFP} in Isabelle. 
Urban, Pitts, and Gabbay \cite{urban2004nominal} also formalized first-order unification in Isabelle. 
A relatively recent formalization for syntactic unification is from Avelar, Galdino, deMoura, and
Ayala-Rincon \cite{avelar2014first} using PVS (Prototype Verification System).

\emph{E}-unification is unification modulo an equational theory.
Dougherty \cite{dougherty2019coq} has verified two algorithms for Boolean 
unification.  Ayala-Rinc{\'o}n \emph{et.\ al.}\ \cite{ayala2022certified} have
verified an AC(Associativity and Commutativity)-Unification algorithm using
PVS. For XOR unification, there are only a
few algorithms and no formalization. 
Tuengethal, Kusters and Turuani  \cite{tuengerthal2006implementing}
mentioned a relatively easy and intuitive way to design such an
algorithm by combining theories such that their overall output
satisfies the XOR properties. 
Guo, Narendran, and Wolfram \cite{guo1996unification} mentioned using Gaussian 
elimination over a Boolean ring to compute unifiers for XOR
unification. %
Liu and Lynch \cite{liu2011efficient}
give several terminating inference rules to solve XOR unification.  %
However, the above papers only give algorithms but not a formalization. Therefore 
in this paper, we decided to do a formalization over their work in Coq so we can be more confidence in the algorithm.
 
\section{Background}

The XOR operator is common operator seen in many protocols; a famous example is the Vernam Cipher\cite[Definition.1.39]{10.5555/548089}:
\begin{equation}
    c_i=m_i\oplus k_i, 1\leq i \leq t.
\end{equation}
where $t$ is the length of the message in digits, $i$ ranges over the digits of the message, $c_i$ is the digit's ciphertext, $m_i$ the message, and $k_i$ the key. 
That is, the Vernam cipher XOR's a message with a key of the same length.
Such a cipher is decrypted by applying the XOR operation a second time with the same key.

We formally state the axioms for XOR, where the signature is $\Sigma=\{\oplus,0\}$:
\begin{itemize}
    \item Associativity: $x\oplus (y\oplus z) = (x\oplus y)\oplus z$
    \item Commutativity: $x\oplus y = y\oplus x$
    \item Unity: $x\oplus 0 = x$
    \item Nilpotency: $x\oplus x = 0$
\end{itemize}

Then, we present the modified rewrite system that is used to compute the unifiers.
This rewrite system \emph{amounts to an algorithm} for XOR Unification, i.e., the algorithm is to apply the first applicable rule, terminating when none applies.
In this paper, we use two sets $\Gamma\|\Lambda$ to keep track of the computation 
progress: $\Gamma$ denotes the unification problem consisting of a set of 
equations $\{S\approx_E^? 0\}$, where $S$ is any term, $\approx_E^?$ is the symbol 
for deciding two terms on both side are the same or not under equational theory E, 
and $0$ is the unit term.
Because of the Nilpotency Axiom, we can move the term from the right-hand side to the 
left-hand side without losing equivalency; i.e. 
$t_1\approx_E^?t_2\rightarrow t_1\oplus t_2\approx_E^? t_2\oplus t_2$ and 
$t_2\oplus t_2 \approx_E 0$. And $\Lambda$ denotes a set of equations in 
solved form. Initially, the unification problems are stored in $\Gamma$, while $\Lambda$ 
remains empty. 
If a system is in normal form regarding these inference rules, then $\Lambda$ is in solved form if the original problem is solvable.

We introduce the two inference rules used in this algorithm:

\textbf{Trivial}: seeks a problem that is already solved and deletes it
\begin{equation}
    \frac{\Gamma\cup\{0\approx_E^?0\}\|\Lambda}{\Gamma\|\Lambda}
\end{equation}

\textbf{Variable Substitution}: seeks a solved form and applies this substitution to the whole system
\begin{equation}
    \frac{\Gamma\cup\{x\oplus S\approx_E^?0\}\|\Lambda}{\sigma\Gamma\|\sigma\Lambda\cup\{x\approx_E^?S\}}
\end{equation}
where $\sigma={x\mapsto S}$ and $x\not\in S$, i.e. the occurs check passes.

In the Coq development, we need to prove this set of inference rules correct. Correct here means it will return an idempotent mgu (most general unifier) of the original problem if it is solvable, and this rewrite system will terminate for all input, see the formal theorem stated in Figure 4. 

\section{Coq Implementation}
This section illustrates the definition of different data structures, the algorithm, and the theorems in Coq. Please note that we only provide the statement of the theorems in this section, as the full proofs have 11,000 lines of code. For the complete development, please refer to our Coq code \cite{CoqCode}. For an introduction to Coq notation, see background material \cite{CoqIntro}.

\subsection{Basic Data structure}
Given that this work only concerns constants, variables, and the XOR operator
($\oplus$), abstract syntax of formulas can be described in the following way in Coq.

    \vspace{-1em}
\begin{figure}[!hbtp]
    \centering
    \begin{lstlisting}
    Definition var := string.
    Inductive term: Type :=
    | C  :  nat -> term #Constant
    | V  :  var -> term #Variable
    | Oplus  : term -> term -> term.
    Definition T0 :term:= C 0. # Short for Constant 0, unit in unity axiom
    Notation "x +' y" := (Oplus x y) (at level 50, left associativity).
    \end{lstlisting}
    
    \vspace{-1em}
    \caption{Term Definition}
    \label{fig:defterm}
\end{figure}

The constructor \texttt{C} takes a natural number, which is a built in data structure from Coq, 
as its input and outputs a constant term, while the constructor \texttt{V} takes a string as input and outputs a variable term. 
The operator $\oplus$ takes two terms as inputs and outputs a nested $\oplus$ term. 
Note that constant \verb|T0| is the unit of XOR.

\begin{figure}[hbt!]
    \centering
    \begin{lstlisting}
    Reserved Notation "x == y" (at level 70).
    Inductive eqv : term -> term -> Prop :=
    | eqvA: forall x y z,    (x +' y) +' z == x +' (y +' z)
    | eqvC: forall x y, x +' y ==  y +' x
    | eqvU: forall x,  T0 +' x  == x
    | eqvN: forall x,  x +' x  == T0
    | eqv_ref: forall x,  x == x  
    | eqv_sym: forall x y,  x == y ->  y == x
    | eqv_trans: forall x y z,  x == y ->  y == z ->  x == z                                 
    | Oplus_compat : forall x x' ,  x == x' -> forall y y' ,
          y == y' -> x +' y ==  x' +' y'
    where "x == y" := (eqv x y).
    \end{lstlisting}
    
    \vspace{-1em}
    \caption{Equivalence Definition}
    
    \vspace{-1em}
    \label{fig:defeqv}
\end{figure}

After introducing the fundamental term representations in Coq, it is necessary to define 
the equivalence relation modulo XOR (shown in Figure 2). In addition to the four axioms of associativity, 
commutativity, unity, and nilpotency, this relation must also satisfy the properties of 
reflexivity, symmetry, and transitivity, as it is an equivalence relation. Since this is
a congruence relation, we also must define a compatibility axiom oplus compat.

\subsection{XOR-Rewrite System}
Processing proofs with nested terms is not a simple job, consider the following two terms:
\vspace{-1em}
\begin{equation}\label{hard-example1}
    z\oplus a\oplus (b \oplus c) \oplus a \oplus (b \oplus c)\oplus z 
\end{equation}
\vspace{-2.1em}
\begin{equation}\label{hard-example2}
    d\oplus (a\oplus e) \oplus ((b\oplus (d \oplus e)) \oplus c) \oplus a \oplus (b \oplus c) 
\end{equation}
Both terms can ultimately be reduced to zero; however, accomplishing this reduction is challenging with nested forms, and the complexity is further increased when utilizing the Coq notation described earlier.

Consequently, an alternative approach was adopted. 
We reduce each term into a list-shaped normal form.
More specifically, we designed two functions: $f_{tlt}()$ and $f_{ltt}()$, to transform a term into a list and back. 
We use the name \texttt{lTerm} to describe this representation, and a predicate $\approx\approx$ for equivalency between \texttt{lTerm}. 
Then, we need to prove that these two predicates capture the same equivalence for these 
two data representations. We state our lemmas below:

\begin{lemma}
    $\forall (t1,t2:term), t1 \approx_{XOR} t2 \leftrightarrow f_{tlt}(t1)\approx\approx f_{tlt}(t2)$
\end{lemma}

\begin{lemma}
    $\forall (tl1,tl2:lTerm), tl1 \approx\approx tl2 \leftrightarrow f_{ltt}(tl1)\approx_{XOR} f_{ltt}(tl2)$
\end{lemma}

Then we designed a rewrite system $f_R()$ such that equivalent \texttt{lTerm}s are syntactically  equal after rewriting. 
In other words, the following theorem must hold true:

\begin{theorem}
    $\forall (tl1,tl2:lTerm), tl1 \approx\approx tl2 \leftrightarrow f_R(tl1)=f_R(tl2)$
\end{theorem}

Once we have the lemmas and theorem in place, we can use the syntactic checker, which checks that both sides are identical, to compute the results for XOR equivalence. 
Note that the other benefit of this 
approach to prove the correctness of unification is that this can be 
easily modified to include homomorphic functions and uninterpreted functions in the future. 
The reason why we developed this representation is because it allowed us to normalize terms and minimize their syntactic complexity which makes the proofs easier.

\subsection{XOR-Unification Algorithm and Correctness}
To set up the final algorithm, we first need to convert the raw input problems to reduced problems (\texttt{lTerm} form), and then perform a fixed number of steps on the reduced problem. If the left-hand side of the problem set is empty after these steps, then the problem is solvable, and the function returns the right-hand side, which is the solved form of the original problem, and then transforms it into a substitution. If the left-hand side is not empty, it means the problem is not solvable and the function returns \texttt{None}.

Here are properties we proved in Coq for correctness: If the algorithm returns \texttt{None} then original unification problem is not solvable. If the algorithm returns some substitution, then this substitution solves the original unification problem.
Specifically, the substitution is the most-general unifier (mgu) of the problem and is idempotent.
\vspace{-0.5em}
\begin{figure}[h!]
    \centering
    \begin{lstlisting}
    Theorem XORUnification_solves:forall(ps:problems)(sb:sub),
      XORUnification ps = Some sb -> solves_problems sb ps.
      
    Theorem XORUnification_mgu:forall(ps:problems)(sb:sub),
      XORUnification ps = Some sb -> mgu_xor sb ps.

    Theorem XORUnification_idpt:forall(ps:problems)(sb:sub),
      XORUnification ps = Some sb -> idempotent sb.
    \end{lstlisting}
    \vspace{-1em}
    \caption{Solution found imply solution, solves, most general and idempotent}
    \vspace{-1em}
    \label{theo:SimplyMGU}
\end{figure}

We also proved that if the problem does not have a solution, then the algorithm will return \texttt{None}.

\vspace{-1em}
\begin{figure}[h!]
    \centering
    \begin{lstlisting}
    Definition problems_unifiable(ps:problems):Prop:=
      exists sb:sub, solves_problems sb ps.
      
    Theorem unifiable_return_sub:forall(ps:problems),
      problems_unifiable ps -> (exists sb:sub, XORUnification ps = Some sb).
      
    Theorem not_unifiable_return_None:forall(ps:problems),
      ~(problems_unifiable ps) -> XORUnification ps = None.
    \end{lstlisting}
    \vspace{-1em}
    \caption{Not solvable implies no solution found}
    \vspace{-1em}
    \label{fig:enter-label}
\end{figure}

To sum up, in this development, we proved that:
If the original unification problem is solvable, then the algorithm will return a
substitution that is a most general unifier and is idempotent. If the original 
unification problem is not solvable then the algorithm will return \texttt{None}.

The Project took 1 person-year. 
This includes time spent learning Coq and iterating on intermediate proof attempts. 
This process included learning the basics of term rewriting, deciding  project scope, implementation, and revision. 
After 1 person-year of work, the Coq implementation has roughly 11,000 lines of code. 
Since most of the proofs are not automated, the complete proof can be checked in Coq in less than 1 second.

In the end, the difficulty of proving soundness and completeness was similar because they rely on the same supporting lemmas. 
The reason why we chose the approach of reducing
terms to \texttt{lTerm}s is because we tried to prove things with nested terms at beginning, but comparing equivalency between two terms is a major challenge as illustrated in Figures 4 and 5.  
This project involved constantly substituting in new terms and comparing two terms to see whether they are solved or whether the equation is balanced. The lack of an efficient decision procedure for equivalence becomes a substantial problem during some specific proofs involving substitution, motivating our approach.
This helped with processing complicated proofs relating to nested terms, because  substitution takes place in \texttt{lTerm} and comparing two terms consists of checking whether their reduced forms are syntactically the same.
This syntactic check is easier on lists.
Moreover, we believe lots of  other equational theories can adopt similar approaches for  reasoning about equivalence. 

Some problems we took a long time to prove are surprisingly not related to the inference 
rules or their correctness. 
Most are about reducing terms to \texttt{lTerm}s.
One key  challenge was building a library for showing intuitive properties about the syntax tree, such as proving that different constructors yield disequal terms. 
For example, \texttt{C} 1 is not equivalent 
to \texttt{V} "x", and \texttt{C} 1 is not equivalent to \texttt{C} 2.

\section{Conclusion and Future Work}
We highlight two areas for future work.
First, in the short term, we propose to extend our Coq formalization with uninterpreted and homomorphic functions.
Second, in the long term, we propose that our formalization of XOR Unification can be used as the basis of a formalized implementation of logic programmnig with XOR, useful as a tool for security protocol analysis.


Supporting uninterpreted and homomorphic functions would require extensions to our  data structures, rules, and proofs.
The data structure changes would be only a few lines of code, and the rules would not be much longer.
However, the proofs would be complicated significantly, as they would need to keep track of sets of uninterpreted and homomorphic functions throughout.
Moreover, the addition of uninterpreted function symbols causes the unification algorithm to become non-deterministic.
The addition of non-determinism increases the conceptual difficulty of the proofs.

In the long term, automated analysis of security exploits using logic programming with XOR is an exciting potential application of this work.
Searching for security exploits in general is an established application of logic programming \cite{zech2019knowledge}.
This suggests logic programming-based approaches may also be worth exploring when analyzing security protocols specifically.
In order to express many security protocols \cite{10.5555/548089} as logic programs, one must support an XOR operator and thus XOR unification.

Security tools are worthy of the strongest possible correctness guarantees, an observation which highlights the potential impact of our work.
Our work provides the first formalization of XOR unification with a machine-checked proof from which guaranteed-correct code can be extracted.
In so doing, we provide a strong foundation for any high-stakes analysis using logic programming with XOR.

\nocite{*}
\bibliographystyle{eptcs}
\bibliography{generic}
\end{document}